# Galilean invariant exchange correlation functionals with quantum memory

Yair Kurzweil and Roi Baer♦

*Department of Physical Chemistry and the Lise Meitner Minerva-Center for Computational Quantum Chemistry, the Hebrew University of Jerusalem, Jerusalem 91904 Israel.*

Today, most application of time-dependent density functional theory (TDDFT) use adiabatic exchange-correlation (XC) potentials that do not take into account non-local temporal effects. Incorporating such "memory" terms into XC potentials is complicated by the constraint that the derived force and torque densities must integrate to zero at every instance. This requirement can be met by deriving the potentials from an XC action that is Galilean invariant (GI). We develop a class of simple but flexible forms for an action that respect these constraints. The basic idea is to formulate the action in terms of the Eularian-Lagrangian transformation (ELT) metric tensor, which is itself GI. The general form of the XC potentials in this class is then derived and the linear response limit is derived as well.

Time dependent density functional theory (TDDFT) [1] is routinely used in many calculations of electronic processes in molecular systems. Almost all applications use "adiabatic" potentials describing an immediate response of the Kohn-Sham potential to the temporal variations of the electron density. The shortcomings of these potentials were studied by several authors[2-5]. Some of the problems are associated with self interaction, an ailment inherited from ground-state density functional theory[6]. Other deficiencies are known or suspected to be associated with the adiabatic assumption. The first attempt to include non-adiabatic effects[7] was based on a simple form of the exchange-correlation (XC) potential in the linear response limit. Studying an exactly solvable system, this form was shown to lead to spurious time-dependent evolution[8]. The failure was traced back to violation of a general rule: the XC force density, derived from the potential, should integrate to zero [9]. Convincing arguments were then presented[10], demonstrating that non-adiabatic effects cannot be easily described within TDDFT and instead a *current density* based theory must be used. Vignale and Kohn [10] gave an expression for the XC potentials applicable for linear response and long wave lengths.

That the total XC force is zero is a valid fact not only in TDDFT but also in TDCDFT. It stems from the basic requirement that the total force on the non-interacting particles must be equal to the total force on the interacting particles. This is so otherwise a different total acceleration results and the two densities or current densities will be at variance. In the interacting system the total (Ehrenfest) force can only result from an external potential: because of Newton's third law the electrons cannot exert a net force upon themselves. In TDCDFT the total force equals the sum of the external force, the Hartree force and the XC force. Since the Hartree force integrates to zero (Newton's third law again) the total XC force do so as well. A similar general argument can be applied to the total torque, showing that the net XC torque must be zero. These requirements then have to be imposed on the approximate XC potentials[9].

The question we deal with in this paper is the how to construct simple approximations to the XC potentials that ensure zero XC force and torque. One way to enforce the zero XC force condition is via the requirement that potentials be derived from a TDCDFT action that is Galilean invariant. The XC action $S[\mathbf{u}]$ is a functional of the electron fluid velocity ($\mathbf{u}(\mathbf{r},t) = \mathbf{j}/n$ where $n(\mathbf{r},t)$ and $\mathbf{j}(\mathbf{r},t)$ are the particle and current densities) defined on a Keldysh contour[11, 12], from which the vector potential $\mathbf{a} = \delta S/\delta \mathbf{u}$ is obtained as a functional derivative. Demanding that it is Galilean invariant means that observers in different frames report the same value of the XC action. Galilean frames can be translationally or rotationally accelerating. In variance in the first case is calle translational invariance (TI) and in the second case, rotational invariance (RI). We discuss this in more detail bellow. Kurzweil and Baer[12] have recently developed a general TDCDFT derived from a TI XC action. Their XC action was however not RI and so did not enforce the zero torque condition. It is the purpose of this paper to further develop the theory along similar lines, to achieve zero XC torque as well. We limit our discussion to as simple a theory as possible, by considering as building blocks only low order derivatives of basic quantities.

As noted above, Galilean invariance of the action means that observers in different Galilean frames report the same value for the XC action. We consider two types of relative motion: translational and rotational. One observer, using "unprimed" coordinates, denotes the current density as $\mathbf{j}(\mathbf{R},t)$ and particle density as $n(\mathbf{R},t)$. A second observer is using primed coordinates. The primed origin is accelerating with respect to that of the unprimed origin where its location is $\mathbf{x}(t)$. A given point in space designated as $\mathbf{R}$ by the first observer and $\mathbf{R}' = \mathbf{R} + \mathbf{x}(t)$ by the second. Here we assume that the axes of the two coordinate systems are parallel, i.e. there is no rotation. Since both observers are studying the same electronic system, the density and velocity functions must be related by:

$$n'(\mathbf{R}',t) = n(\mathbf{R},t) = n(\mathbf{R}' - \mathbf{x}(t),t)$$
$$\mathbf{u}'(\mathbf{R}',t) = \mathbf{u}(\mathbf{R},t) + \dot{\mathbf{x}}(t) = \mathbf{u}(\mathbf{R}' - \mathbf{x}(t),t) + \dot{\mathbf{x}}(t) \quad (1.1)$$

In ref. [12] we showed that in order to obtain zero XC force, we demand translational invariance i.e. $S[\mathbf{u}] = S[\mathbf{u}']$.

Zero total XC-torque is guaranteed when the XC action is

---

♦ Corresponding author: FAX: +972-2-6513742, roi.baer@ huji.ac.il

RI, $S[\mathbf{u}] = S[\mathbf{u}'']$ where the double-primed quantities are related to the coordinate system of a third observer whose axis is rotating around the common origin. At time $t$ the point $\mathbf{R}$ in space will be labeled by this observer as: $\mathbf{R}'' = M(t)\mathbf{R}$ where $M(t)$ is some instantaneous orthogonal matrix (with unit determinant) describing the rotated axes (for convenience, we assume that $M \equiv 1$ when $t = 0$). The density and velocity fields as defined by this third observer are:

$$n''(\mathbf{R}'',t) = n(\mathbf{R},t) = n(M(t)^{-1}\mathbf{R}'',t)$$
$$\mathbf{u}''(\mathbf{R}'',t) = M(t)\mathbf{u}(\mathbf{R},t) + \dot{M}(t)\mathbf{R} \quad (1.2)$$
$$= M(t)\mathbf{u}(M(t)^{-1}\mathbf{R}'',t) + \dot{M}(t)M(t)^{-1}\mathbf{R}''$$

We want to describe now a method for generating GI actions. The way we follow is to identify GI quantities and write the action in terms of them. What are the simply accessible GI quantities? We follow previous works [8, 12, 13] and consider the Lagrangian coordinates, $\mathbf{R}(\mathbf{r},t)$ defined by:

$$\dot{\mathbf{R}}(\mathbf{r},t) = \mathbf{u}(\mathbf{R}(\mathbf{r},t),t) \qquad \mathbf{R}(\mathbf{r},0) = \mathbf{r} \quad (1.3)$$

$\mathbf{R}(\mathbf{r},t)$ is the position at time $t$ of a fluid element originating at a point labeled $\mathbf{r}$; in other words, $\mathbf{R}(\mathbf{r},t)$ is the trajectory of the fluid element $\mathbf{r}$. The coordinate $\mathbf{r}$ can be viewed as a Eularian coordinate, so $\mathbf{R}(\mathbf{r},t)$ is the Eularian-Lagrangian transformation (ELT). Inventing memory functionals in the Lagrangian frame is easier because local memory is naturally described *within* a fluid element.

It can be readily checked that that the Lagrangian density $N(\mathbf{r},t) = n(\mathbf{R}(\mathbf{r},t),t)$ is in fact GI, i.e. it is invariant with respect to both linear and rotational accelerating observers. Consider first accelerations. We assume both observers label the different fluid elements in the same way (i.e. their axes coincide at $t = 0$). Thus: $\mathbf{R}'(\mathbf{r},t) = \mathbf{R}(\mathbf{r},t) + \mathbf{x}(t)$ and from (1.1):

$$N'(\mathbf{r},t) = n'(\mathbf{R}'(\mathbf{r},t),t) = n'(\mathbf{R}(\mathbf{r},t) + \mathbf{x}(t),t)$$
$$= n(\mathbf{R}(\mathbf{r},t),t) = N(\mathbf{r},t). \quad (1.4)$$

Here and henceforth we use the notation $\partial_i \equiv \partial/\partial r_i$, $i = 1,2,3$. A rotating observer with the same labeling convention sees $\mathbf{R}''(\mathbf{r},t) = M(t)\mathbf{R}(\mathbf{r},t)$, so from (1.2):

$$N''(\mathbf{r},t) = n''(\mathbf{R}''(\mathbf{r},t),t) = n(\mathbf{R}(\mathbf{r},t),t) = N(\mathbf{r},t), \quad (1.5)$$

Eqs. (1.4) and (1.5) show that $N(\mathbf{r},t)$ is indeed GI so a simple form for the action functional can be immediately written down as $S^{(1)}[\mathbf{u}] = s_1[N[\mathbf{u}]]$. Looking for a more general yet still simple form, we now consider the Jacobian matrix of the ELT:

$$\Im_{ij} = \partial_j R_i(\mathbf{r},t) \quad (1.6)$$

This matrix is TI, as can be straightforwardly verified[12]. However, $\Im$ is not RI. Indeed, the following transformation, derived from the definition of the rotation, $\mathbf{R}'' = M(t)\mathbf{R}$, must hold:

$$\Im''(\mathbf{r},t) = M(t)\Im(\mathbf{r},t) \quad (1.7)$$

While $\Im$ is not GI, its determinant is: since $\det \Im'' = \det M \det \Im$ and $\det M = 1$. One can then suggest that $S^{(2)}[\mathbf{u}] = s_2[\det \Im[\mathbf{u}]]$. Comparing with $S^{(1)}$ though, we find $S^{(2)}$ contains nothing new! This is because the function $N(\mathbf{r},t)$ is directly related to the Jacobian determinant. Indeed, the number of particles in a fluid element must be constant so $n(\mathbf{R}(\mathbf{r},t),t)d^3R = n(\mathbf{r},0)d^3r$, and thus:

$$J(\mathbf{r},t)^{-1} \equiv \left|\det[\Im(\mathbf{r},t)]\right|^{-1} = N(\mathbf{r},t)/n_0(\mathbf{r}), \quad (1.8)$$

where $n_0(\mathbf{r}) = n(\mathbf{r},0)$. Thus, the functional $s_1$ can also be thought of as a functional of $\det[\Im]$.

Our first attempt to introduce an action in terms of $\Im$ yielded nothing new. Let is return to Eq. (1.7) and search for another invariant quantity. This leads us to consider the $3 \times 3$ symmetric positive-definite ELT metric tensor:

$$g(\mathbf{r},t) = \Im(\mathbf{r},t)^T \Im(\mathbf{r},t). \quad (1.9)$$

It is immediately obvious from Eq. (1.7) and the orthogonality of $M(t)$ that $g(\mathbf{r},t) = g''(\mathbf{r},t)$. Thus $g$ is RI. It is also TI[12]. Thus, it is GI. The tensor $g$ essentially tells us how to compute the distance $dS$ between two infinitesimally adjacent fluid elements, at $\mathbf{r}$ and $\mathbf{r} + d\mathbf{r}$:

$$dS^2 = (\mathbf{R}(\mathbf{r}+d\mathbf{r},t) - \mathbf{R}(\mathbf{r},t))^2 = d\mathbf{r}^T \cdot g \cdot d\mathbf{r}. \quad (1.10)$$

Now it is clear why $g$ is GI: any two observers will agree upon the distance between any two electron-fluid parcels.

The metric tensor is thus a natural quantity on which the action can defined. Thus, we consider the following class of metric-tensor actions:

$$S[n_0, \mathbf{u}] = s[n_0, g[\mathbf{u}]]. \quad (1.11)$$

Here, $n_0$ is the initial ground-state density (assuming the system starts from its ground-state). It is comforting to note that in view of Eq. (1.8) and the fact that $|\det \Im| = \sqrt{\det g}$, this form includes $S^{(1)}$ as a special case.

The potential derived from (1.11) is obtained from a chain-rule functional derivation. Defining the symmetric tensor

$$Q_{ji}(\mathbf{r}',t') \equiv 2\frac{\delta s[n_0,g]}{\delta g_{ji}(\mathbf{r}',t')}, \quad (1.12)$$

where the factor of 2 appears for later convenience. The form of the vector potential is obtained by considering the action change resulting from perturbing the velocity field at time $t$ and position $\mathbf{R} \equiv \mathbf{R}(\mathbf{r},t)$:

$$a_k(\mathbf{R},t) = \frac{1}{2}\iint Q_{ji}(\mathbf{r}',t')\frac{\delta g_{ji}(\mathbf{r}',t')}{\delta u_k(\mathbf{R},t)}dt'd^3r', \quad (1.13)$$

(here we use the convention that repeated indices are summed over). We note that the integration on time here is actually an integration over the Keldysh contour[11], described fully in ref [12]. The derivative is given by:

$$\frac{\delta g_{ji}(x')}{\delta u_k(X)} = \left[\Im_{li}(x')\partial'_j + \Im_{lj}(x')\partial'_i\right]G_{lk}(x';X) \quad (1.14)$$

Where $x' \equiv (\mathbf{r}',t')$, $X \equiv (\mathbf{R},t)$ and $G_{ij}$ is derived in ref. [12], given by:

$$G_{lk}(x';X) = \left[\Im(\mathbf{r}',t')\Im(\mathbf{r}',t)^{-1}\right]_{lk}\theta(t'-t)\delta(\mathbf{R}(\mathbf{r}',t)-\mathbf{R}) \quad (1.15)$$

Using Eqs. (1.14) and (1.15) in (1.13), we find, integrating by parts:

$$a_k(\mathbf{R},t) = -\iint \partial_i\left[Q_{ij}(\mathbf{r}',t')\Im_{lj}(\mathbf{r}',t')\right]G_{lk}(\mathbf{r}',t';\mathbf{R},t)d^3r'dt' \quad (1.16)$$

leading to the following general form vector potential:

$$\mathbf{a}(\mathbf{R}(\mathbf{r},t),t) = J(\mathbf{r},t)^{-1}\Im(\mathbf{r},t)^{-1}\mathbf{A}(\mathbf{r},t) \quad (1.17)$$

Where:

$$A_m(\mathbf{r},t) = \int_0^t \Im(\mathbf{r},t')_{ml}^T \partial_i\left[\Im(\mathbf{r},t')Q(\mathbf{r},t')\right]_{li} dt' \quad (1.18)$$

Is the "Lagrangian" vector potential. An explicit derivation of equation (1.18) shows that it is the time integral $\int_\infty^t$ instead of $\int_0^t$ which appears. However, had we made the development on a Keldysh contour the correct from of the integral (1.17) would have resulted. The procedure was demonstrated in ref. [12].

Eqs. (1.11), (1.17) and (1.18) are the central result of this paper, resulting in a general form for a potential which yields zero force and torque. This general form should find useful application in cases where the electronic systems interact with strong fields.

We would like to compare our results with previous work on TDCDFT potentials in the linear response regime[10, 14]. For this purpose, consider Eqs. (1.17) and (1.18) developed up to first order quantities, linear in the perturbation: $\mathbf{R} \to \mathbf{r} + \mathbf{R}_1$, $\Im \to 1+\Delta$, $J \to 1+tr\Delta$ and $Q \to q(\mathbf{r}) + \theta(\mathbf{r},t)$ where the $q(\mathbf{r})$ is a zeroth-order term and $\theta(\mathbf{r},t)$ is the first order term, given by the following expression: $\theta_{mi}(\mathbf{r},t) = \int d^3r' \int_0^t \Theta_{mi}^{kl}(\mathbf{r},\mathbf{r}'t-s)\Delta_{kl}(\mathbf{r}',s)ds$. Calculating the vector potential to zero and first orders we obtain:

$$\mathbf{a}_0 + \mathbf{a}_1 = \mathbf{A}_0 + \mathbf{A}_1 - (\Delta + [tr\Delta]I)\mathbf{A}_0 \quad (1.19)$$

Where the zero order term $\mathbf{A}_0$ is given by:

$$(A_0)_m(\mathbf{r},t) = t\partial_i q_{mi}(\mathbf{r}) \quad (1.20)$$

After some algebra, moving into the frequency domain, replacing $\Delta_{ji}(\mathbf{r},\omega)$ with $(i\omega)^{-1}\partial_i u_j(\mathbf{r},\omega)$, we find a first-order response given by the following expression:

$$[a_1]_m(\mathbf{r},\omega) = \frac{1}{\omega^2}\Big\{[\partial_m u_l + \partial_l u_m]\partial_i q_{li} + q_{li}\partial_i\partial_l u_m \\
+ \partial_i\left(\int \Theta_{mi}^{kl}(\mathbf{r},\mathbf{r}',\omega)\partial'_l u_k(\mathbf{r}',\omega)d^3r'\right)\Big\} \quad (1.21) \\
+ \frac{\partial}{\partial\omega}\left[\frac{1}{\omega}(\partial_l u_m + \partial_k u_k\delta_{ml})\right]\partial_i q_{li}$$

It is possible to show that this form is compatible with the VK result for a HEG[10]:

$$\mathbf{a}(\mathbf{r},\omega) = \frac{1}{\omega^2}\{f_T\nabla\times[\nabla\times\mathbf{u}] - f_L\nabla[\nabla\cdot\mathbf{u}]\} \quad (1.22)$$

where $f_L$ and $f_T$ are the HEG response kernels. This is achieved by taking $\Theta_{mi}^{kl}(\mathbf{r},\mathbf{r}',\omega) = \tilde{\Theta}_{mi}^{kl}(\omega)\delta(\mathbf{r}-\mathbf{r}')$ and have the kernels $q_{li}$ and $\tilde{\Theta}_{mi}^{kl}$ obey:

$$\tilde{\Theta}_{mi}^{kl} + q_{li}\delta_{mk} = (\delta_{ml}\delta_{ik} - \delta_{mk}\delta_{il})f_T - f_L\delta_{im}\delta_{lk} \quad (1.23)$$

Note that for HEG $q = q[n_0] \Rightarrow \partial_i q = 0$.

Following the general ideas of Ref. [12] for constructing an action beyond linear response, one can choose $s[g] = \int g(r,t)\Theta(N(r,t'),t-t')g(r,t')d^3rdtdt'$ and obtain through Eq. (1.17) and (1.18) the appropriate potentials. These are compatible with the linear response of the homogeneous electron gas (by choosing $\Theta$ compatible with (1.23)) while yielding zero force and torque beyond the linear response regime. When additional information on the response of an inhomogeneous gas becomes available this can be built into the kernel $\Theta_{mi}^{kl}$.

Summarizing the present work, we have developed a form for an action which yields potentials that are consistent with the zero force and zero torque condition. The action is based on the ELT metric tensor $g$. The metric tensor is also an important quantity when the Kohn-Sham equations are presented in a Lagrangian system[15]. The metric $g$ is therefore emerging as an overall important quantity in any non-adiabatic TDCDFT scheme. The form of the action, Eq. (1.11), may also include dependence on $\dot{g}$, $\ddot{g}$ etc. In this case the functional derivatives $Q$ become differential operators with respect to time. More elaborate functionals can be built following similar ideas at the expense that additional gradients are introduced (the metric tensor already leads to potentials that use up to second gradi-ents of the ELT $\mathbf{R}(\mathbf{r},t)$).

**Acknowledgements** We gratefully acknowledge the support of the German Israel Foundation.